\documentclass[aps,pra,twocolumn,groupedaddress,showpacs,showkeys]{revtex4-1}
\usepackage{nicefrac}
\usepackage{amssymb}
\usepackage{amsmath}
\usepackage{bbold}
\usepackage{graphicx}
\usepackage{rotating}
\usepackage{multirow}
\usepackage{color}
\usepackage{pifont}
\usepackage{array}

\newcommand{\ket}[1]{\left\vert #1 \right\rangle}
\newcommand{\bra}[1]{\left\langle #1 \right\vert}

\usepackage{hyperref}
\begin{document}


\title{Quantum Error Correction with Mixed Ancilla Qubits}


\author{Ben Criger}
\affiliation{Institute for Quantum Computing}
\affiliation{Department of Physics and Astronomy, University of Waterloo}
\email[dcriger@iqc.ca]{}
\author{Osama Moussa}
\affiliation{Institute for Quantum Computing}
\affiliation{Department of Physics and Astronomy, University of Waterloo}
\author{Raymond Laflamme}
\affiliation{Institute for Quantum Computing, University of Waterloo}
\affiliation{Perimeter Institute for Theoretical Physics, Waterloo, Ontario}
\affiliation{Department of Physics and Astronomy, University of Waterloo}


\date{\today}

\begin{abstract}
Most quantum error correcting codes are predicated on the assumption that there exists a reservoir of qubits in the state $\ket{0}$, which can be used as ancilla qubits to prepare multi-qubit logical states. In this report, we examine the consequences of relaxing this assumption, and propose a method to increase the fidelity produced by a given code when the ancilla qubits are initialized in mixed states, using the same number of qubits, at most doubling the number of gates when the recovery operation would already be implemented. The procedure implemented consists of altering the encoding operator to include the inverse of the unitary operation used to correct detected errors after decoding. This augmentation will be especially useful in quantum computing architectures that do not possess projective measurement, such as solid state NMRQIP.
\end{abstract}

\maketitle

\section{Introduction}
Quantum error correction~\cite{PhysRevA.54.1098,PhysRevLett.77.793,PhysRevA.55.900} is the practice of using a large number of error-prone qubits to encode a small amount of information in such a way as to protect the logical state, at least partially, from errors which would otherwise affect it. In general, the preservation of a one-qubit state using an error-correcting code can be divided into four operations. \textbf{1.}~The one-qubit input state $\alpha \ket{0} + \beta \ket{1}$ is attached to an ancilla and a unitary operation rotates the state into the final encoded state $\alpha \ket{\bar 0} + \beta \ket{\bar 1}$, where $\left \lbrace \ket{\bar 0},\,\ket{\bar 1} \right \rbrace$ are a set of orthogonal states in the larger Hilbert space.
\textbf{2.}~Each qubit in the encoded state is subjected to the random error process that the code is designed to correct.
\textbf{3.}~The inverse of the encoding unitary is applied. The density matrix for the state will contain terms proportional to $ U\ket{\psi}\!\bra{\psi} U^{\dagger} \otimes \ket{s}\!\bra{s}$, where $\ket{\psi}$ is the original state and $s$ is a classical $n-1$-bit string, the \textit{syndrome} of the error $U$.  
\textbf{4.}~A unitary, controlled on the syndrome qubits, inverts the unitary in the terms described above, producing a final state which has greater fidelity to the input state than the state resulting from unencoded transmission. 

The function of a quantum error correcting code is to divert entropy accrued during transmission to the ancilla qubits. Thus, it is often assumed that the qubits which comprise the ancilla are initialized in the state $\ket{0}$, or a state with negligible entropy. However, this assumption is often violated in practice. For example, let us consider the state of the ancilla immediately after either procedure in Figure \ref{fig:3bitdirtycode} has been performed. This increases the entropy of the ancilla state, which must be `refreshed' to $\ket{00}$ in order for the code to be used again. If the operation which accomplishes this is imperfect, the ancilla will retain some of the entropy it gained during error correction.

In addition, many quantum computing architectures exist in which qubits equilibrate into Boltzmann distributions. Consider, for example, low-temperature solid-state ESR~\cite{Gershenfeld17011997,2007arXiv0710.1447B,Simmons2011}, where the population of the ground state of an electron spin is $\sim \nicefrac{3}{4}$ at 4.2 Kelvin and 7 Tesla. Throughout the remainder of this report, we treat the initial state as the result of an error process which occurs before the encoding operation.

In the following sections, we detail the error map that produces the ancilla noise we consider, and describe an augmentation to error-correcting codes which prevents some of the deleterious effects of this initialization error. We proceed to test this augmentation on two widely-studied error correction codes, correcting bit-flip and depolarization. We conclude by examining the effects of augmentation on a concatenated code. 
\subsection{Initialization Error}
In this report, we examine the consequences of attempting error correction using mixed ancillary qubits, each in the state
\begin{equation}
\rho_{q} = \left[ \begin{array}{cc}
1-\nicefrac{q}{2} & 0 \\ 0 & \nicefrac{q}{2}
\end{array} \right].
\end{equation}
In order to study the effect of this noise, we wish to model it as the result of an error process. One error process which takes the state $\ket{0}$ to the distribution above is the bit-flip channel described below,
\begin{flalign}
\Lambda=\left \lbrace \sqrt{1-\frac{q}{2}}\hat I,\,\sqrt{\frac{q}{2}}\hat X\right \rbrace,
\end{flalign}
where the channel is given in the Kraus representation $\Lambda \equiv \left \lbrace \Lambda_j \right \rbrace,\,\
\Lambda(\rho)=\sum_j \Lambda_j \rho \Lambda_j^{\dagger} ,\, \sum_j \Lambda_j^{\dagger} \Lambda_j = \hat I$~\cite{citeulike:541803}. The initial state on the ancillae is $\Lambda^{\otimes n} \left[\ket{0}\!\bra{0}^{\otimes n}\right]$. This additional error limits the ability of the ancilla to absorb entropy. As a result, any error-correcting code with ancilla qubits being maximally mixed, with $q=1$, will not be useful in an error-correcting code. 

\section{Channel Fidelity}
In order to study the effects of initialization noise, we also require a quantitative criterion to determine which quantum error correction protocols are useful and which are not. We evaluate the utility of a given protocol using the channel fidelity, a special case of Schumacher's entanglement fidelity~\cite{PhysRevA.54.2614}:
\begin{flalign}
F_C(\Lambda) &= \bra{\Omega} (\Lambda \otimes \hat I) \left[ \ket{\Omega} \bra{\Omega} \right] \ket{\Omega} \nonumber \\ &= \dfrac{1}{4^n}\sum_{k} \vert \textrm{Tr}(\Lambda_k) \vert^2
\end{flalign}
where  $\ket{\Omega} = \dfrac{1}{\sqrt{2^n}}\sum_{j \in \lbrace 0,\, 1\rbrace}^{n} \ket{j} \otimes \ket{j}$ and $\left \lbrace \Lambda_k \right \rbrace$ are the Kraus operators for the channel $\Lambda$. The channel fidelity is a measure of the average fidelity of the output state with the input state~\cite{PhysRevA.60.1888}. The utility of a quantum error correcting code will be decided based on whether its channel fidelity for an input state exceeds the channel fidelity of unencoded transmission through the error channel.

\section{Augmented Error Correction}
\begin{figure}[h!]
\centering
\includegraphics[scale=0.5]{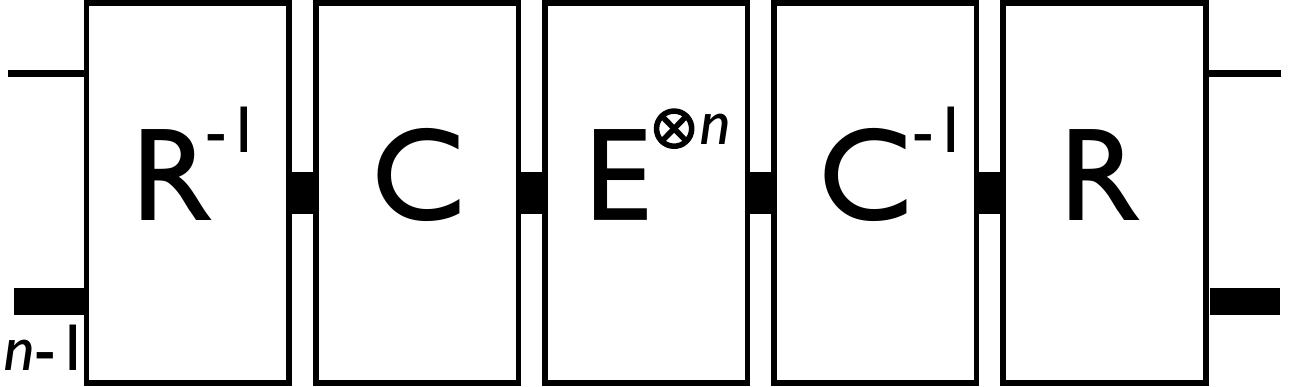}
\caption{(Colour on-line) In order to augment an error correction code on $n$ qubits, the recovery operator is inverted and implemented before encoding. This eliminates faults caused solely by false syndromes. Here, $R$ is the recovery operator, $C$ is the encoding operator and $E$ is the error map. }
\label{fig:genericdirtycode}
\end{figure}

\begin{figure}[h!]
\centering
\includegraphics[scale=0.5]{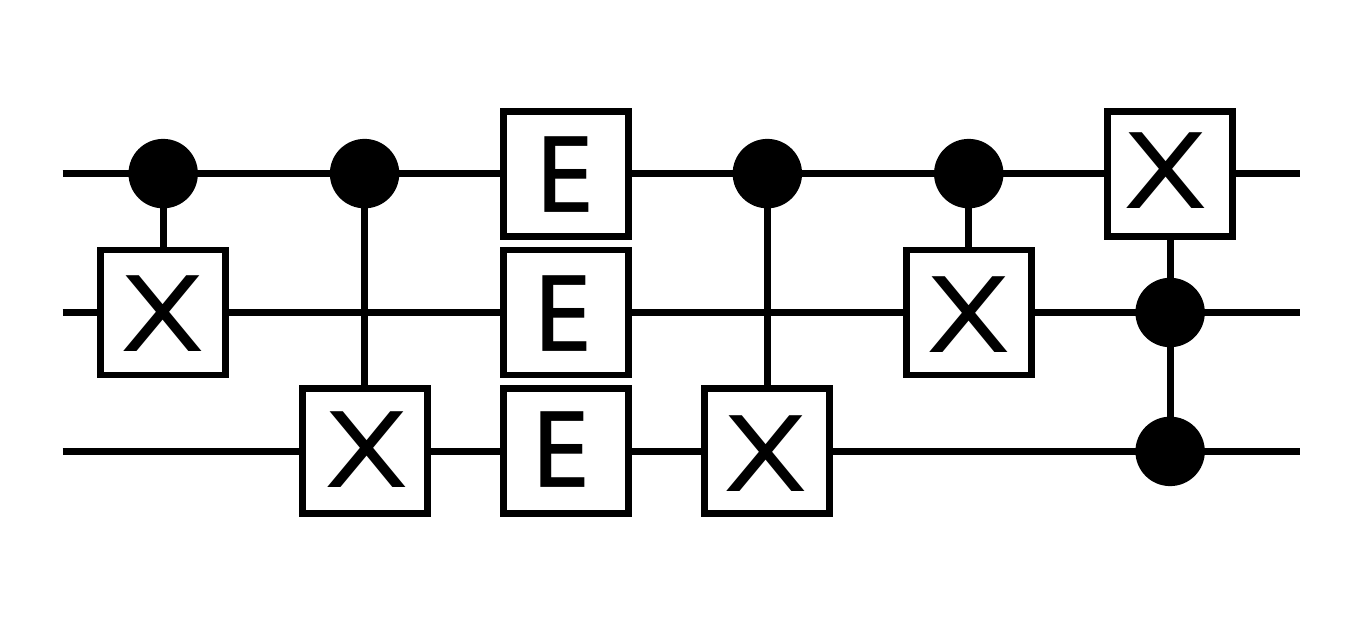}
\includegraphics[scale=0.5]{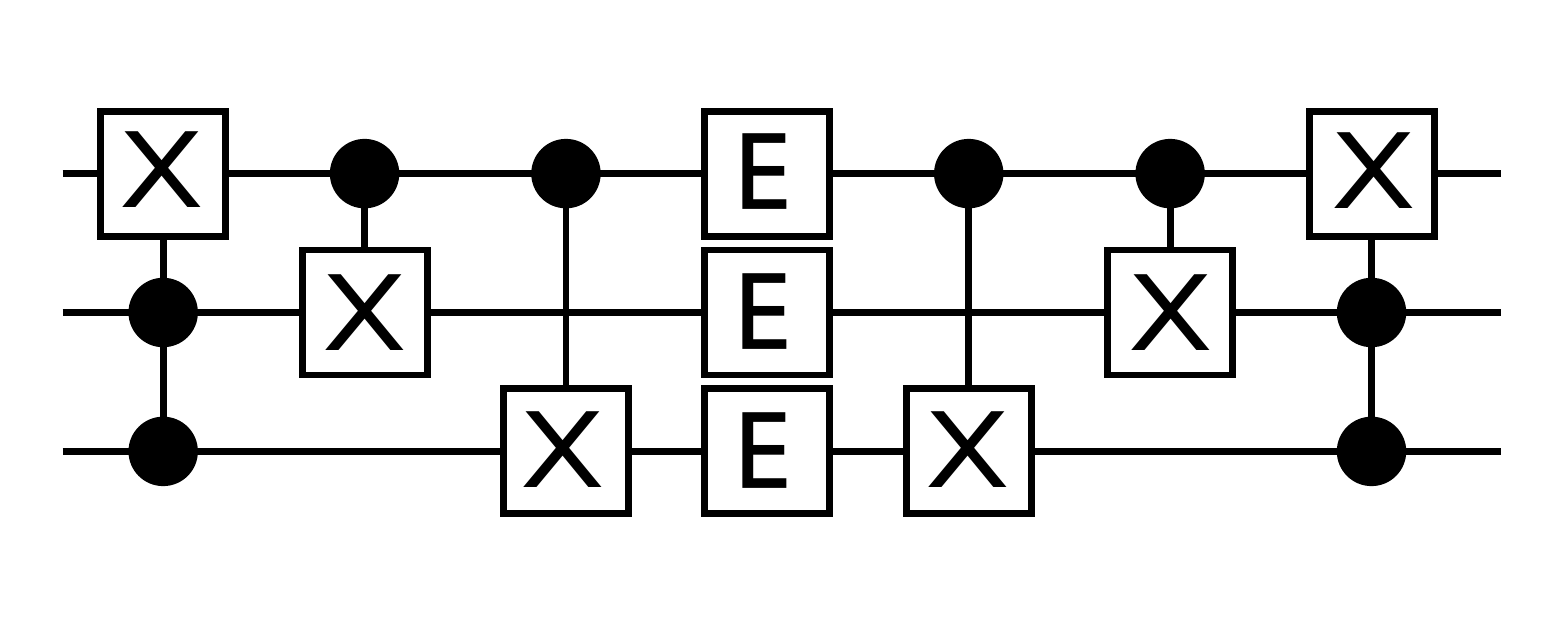}
\caption{(Colour on-line) The traditional 3-qubit code to correct bit-flip errors (above), augmented to provide increased fidelity (below), in the case where each ancilla qubit is subject to the initialization noise map discussed above. The map $E$ in this example is the bit-flip map $\left \lbrace \sqrt{1-p} \hat I,\, \sqrt{p} \hat X \right \rbrace$. The augmentation consists of implementing the Toffoli used to correct detected errors before the standard encoding procedure takes place. This improves the overall fidelity by ensuring that, if no error occurs, the encoded state remains unaltered by false syndromes.}
\label{fig:3bitdirtycode}
\end{figure}

When subjected to initialization error, the controlled operations in the correction stage can introduce new errors into the output state, since the syndrome has been altered by the initialization error. A new code can be created to mitigate this error by implementing the inverse of the correction operation before the encoding unitary. This is shown in Figures \ref{fig:genericdirtycode} and \ref{fig:3bitdirtycode}. 

The augmented three-qubit code in Figure \ref{fig:3bitdirtycode} can be shown to satisfy a numerically-derived upper bound for the channel fidelity using an arbitrary CPTP/unital channel for encoding. To derive this bound, the optimization of channel fidelity is posed as a semi-definite program~\cite{reimpellthesis}, optimizing over the encoding channel, which is linearly constrained to be both CPTP and unital. The numerical search for optimal encoders is the origin of the augmentation in this report. 

\begin{figure}[ht]
\centering
\includegraphics[scale=1]{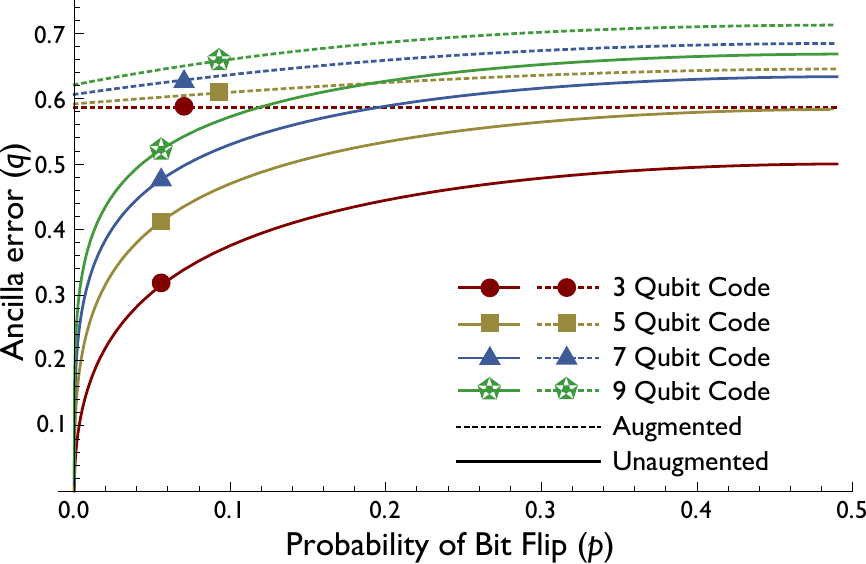}
\caption{(Colour on-line) The initialization error for which an error correcting code can give a channel fidelity $\geq 1-p$. This is shown for four repetition codes correcting bit-flip errors. Note that the tolerable error for small values of $p$, the parameter describing the main bit-flip channel, approaches 0 rapidly for unaugmented codes. By contrast, augmentation provides a high tolerable $q$ for every value of $p$.}
\label{fig:qtol}
\end{figure}

The advantages of this augmentation are:
\textbf{1.}~The fidelity of the augmented codes will always exceed or equal that of the unaugmented codes, since the augmentation corrects additional errors left uncorrected by the unaugmented codes without altering the function of the error correcting code for pure ancillae.
\textbf{2.}~The augmented code is especially useful in implementations where the main error parameter $p$ can be constrained, and the initialization parameter $q$ cannot. For example, when the error parameter during a storage operation is time-dependent, reducing the storage time reduces the error parameter. This is true for any error channel, and any number of ancilla qubits, since the inverse recovery operator prevents faults in the case where the error map acts trivially. Therefore, when $p$ can be diminished to arbitrary size, the augmented code allows arbitrarily high fidelity, where the unaugmented code does not.
\textbf{3.}~Augmented codes provide increased fidelity at higher $q$ than unaugmented codes. A code (whose implementation will be denoted $\Theta$) is \textit{useful} if, for an error channel $\Lambda$, $F_C(\Theta)\geq F_C(\Lambda)$.To illustrate this, we plot the tolerable $q$ in Figures \ref{fig:qtol}, \ref{fig:qtoldep} and \ref{fig:concat_tol_q}.

Furthermore, note that this procedure increases the gate complexity of the code by at most a factor of 2, since the augmenting unitary is already required for the code to function. We conclude that this augmentation will be useful in a variety of circumstances, and in the following sections, we examine examples of this strategy used to counter two common error processes; bit flip and depolarization.  
\section{Bit Flip} 
In order to correct Pauli-$\hat X$ (bit flip) errors, we encode the state we wish to preserve into the 2-dimensional subspace of an $n$-qubit ($2^n$-dimensional) register having maximum distinguishability under bit flip; $\lbrace \ket{0}^{\otimes n}, \ket{1}^{\otimes n} \rbrace$. In order to correct $t^{\textrm{th}}$-order bit flip errors, $2t+1$ qubits are required. Here, we analyse 3-, 5-, 7- and 9-qubit repetition codes to counter bit flip errors, with and without augmentation. Each fidelity is expressed as a polynomial in $p$, $F_C=\sum_k c_k p^k$, the $c_0,\,c_1$ are shown in Table \ref{table:flipcodes}.
\begin{table}
\centering
\setlength{\extrarowheight}{1.5pt}
\begin{tabular}{|c|c|l|l|}
\cline{1-4}
&\# Qubits  & $c_0$ & $c_1$  \\
\cline{1-4}
\multirow{4}{*}{\begin{sideways}{\tiny Unaugmented}\end{sideways}}& 3  & $1-\nicefrac{1}{4}q^2$ &  $-2 q+\nicefrac{3 }{2}q^2$ \\
& 5  & $1-\nicefrac{1}{2}q^3+\ldots$& $-\nicefrac{9}{2} q^2+6 q^3-\ldots$ \\
& 7  & $ 1-\frac{15}{16} q^4+\ldots$& $-10q^3+\ldots$ \\
& 9  & $1-\nicefrac{7}{4} q^5+\ldots$& $-\nicefrac{175}{8} q^4 -\ldots$ \\
\cline{1-4}
\multirow{4}{*}{\begin{sideways}{\tiny Augmented}\end{sideways}}& 3  & 1& $-2 q+\nicefrac{1}{2}q^2$  \\
& 5  & 1& $-\nicefrac{9}{2} q^2+3 q^3+\ldots$ \\
&7  & 1& $-\nicefrac{5}{16} q^3 + \ldots$ \\
&9  & 1& $-\nicefrac{175}{8} q^4-\ldots$ \\
\cline{1-4}
\end{tabular}
\caption{Fidelity coefficients for four repetition codes, correcting bit flip. Each fidelity is expressed as a polynomial in $p$, $F_C=\sum_k c_k p^k$, the $c_0,\,c_1$ are shown. Note that, for the augmented codes, $c_0=1$, indicating that the contribution to the error term due solely to the mixed ancilla has been eliminated.}
\label{table:flipcodes}
\end{table}

We conclude by noting that this behaviour can be trivially extended to codes that correct any channel of the form $\left \lbrace \sqrt{1-p} \hat I,\,\sqrt{p}UXU^{\dagger} \right \rbrace$.
\section{Depolarization}
It is important, in order to ensure that augmented error correction codes are widely useful, to examine the performance of such codes correcting depolarization, an error process to which all error processes can be reduced~\cite{PhysRevA.80.012304}. Depolarization is a channel which consists of the following Kraus map:
\begin{equation}
\Gamma = \left \lbrace \sqrt{1-\dfrac{3p}{4}}\hat I, \sqrt{\dfrac{p}{4}} \hat X, \sqrt{\dfrac{p}{4}} \hat Y, \sqrt{\dfrac{p}{4}} \hat Z \right \rbrace
\end{equation}
We find channel fidelities for an augmented 5-qubit code versus depolarization, and an unaugmented code~\cite{PhysRevLett.77.198,PhysRevA.54.3824}. 

Here, the optimization of the channel fidelity has not been posed as a semi-definite problem. Instead, we have assumed that a unitary will be appended to the encoder which consists of $2^{n-1}$ single-qubit unitaries, each controlled on a unique binary string on the ancilla. This reduces the size of the optimization problem from $4^n$ to $3 \cdot 2^{n-1}$, each single-qubit unitary having 3 free parameters. We observe that the optimal unitary is the inverse of the correcting operation.

\begin{figure}[ht]
\centering
\includegraphics[scale=1]{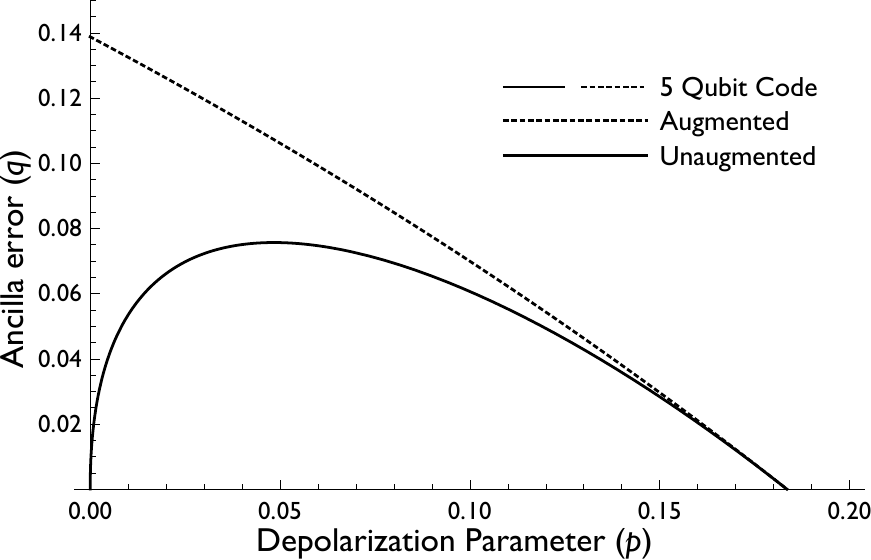}
\caption{The initialization error for which an error correcting code can give a channel fidelity $\geq 1-\dfrac{3}{4}p$. This is shown for the perfect 5-qubit code. Note that the behaviour of this code is qualitatively different, having 0 tolerable initialization for $p\sim 0.18$. The ability of the augmented code to provide  finite tolerable $q$ at $p=0$ is preserved.}
\label{fig:qtoldep}
\end{figure}

We present the polynomial coefficients for the fidelity, as described in Table \ref{table:depcodes}.
\begin{table}
\centering
\setlength{\extrarowheight}{1.5pt}
\begin{tabular}{r|l|l|}
\cline{2-3} 
 & $c_0$ & $c_1$ \\
  \cline{2-3}
 Unaugmented &$1 - \nicefrac{3}{2}q^2 + q^3 - \ldots$ & $-6 q+\nicefrac{21 }{2}q^2-\nicefrac{11 }{2}q^3 + \ldots$\\
 \cline{2-3}
Augmented & 1 & $-6 q+\nicefrac{9 }{2}q^2-\nicefrac{3}{2} q^3+\ldots$\\
\cline{2-3}
\end{tabular}
\caption{Fidelity coefficients for the 5-qubit perfect code, correcting depolarization. Each fidelity is expressed as a polynomial in $p$, $F_C=\sum_k c_k p^k$, the $c_0,\,c_1$ are shown. Coefficients for the unaugmented code are above, those for the augmented code below. Note that, for the augmented codes, $c_0=1$, indicating that the contribution to the error term due solely to the mixed ancilla has been eliminated.}
\label{table:depcodes}
\end{table}
Here, we see that the $p$-independent terms are eliminated, but the term linear in $p$ remains. We continue, showing the tolerable initialization noise levels for codes that counter depolarization errors in Figure \ref{fig:qtoldep}
\section{Concatenation}
\begin{figure}[ht]
\includegraphics[scale=1]{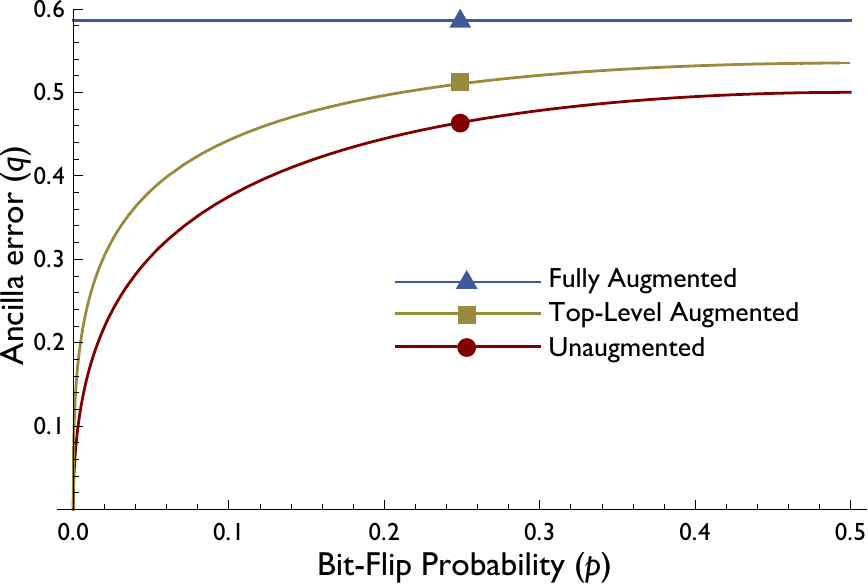}
\caption{(Colour on-line) The tolerable initialization noise for different concatenated codes. For the top-level concatenation, the encoder used in the top of Figure \ref{fig:3bitdirtycode} has been replaced with the encoder used in the bottom of Figure \ref{fig:3bitdirtycode}. For the full concatenation, all the encoding circuits used are augmented, as in Figure \ref{fig:3bitdirtycode}. Note that the bottom curve (for the unaugmented concatenated code) is identical to the tolerable initialization noise for the 3-bit error correcting code when left unconcatenated, shown in Figure \ref{fig:qtol}. Also, the tolerable $q$ for the fully augmented code is $2-\sqrt{2}$, identical to the augmented 3-qubit code.} 
\label{fig:concat_tol_q}
\end{figure}

It is useful to examine the effect of augmentation on a two-level concatenated code, in order to determine the benefits of augmentation at each level. Below, we examine the effect of augmentation on the concatenated 3-qubit code. With bit-flip probability $p$ and initialization error $q$ as defined above, the channel fidelity for unaugmented, top-level augmented, and fully augmented codes are shown in Table \ref{table:concatfids} 
\begin{table}
\centering
\setlength{\extrarowheight}{1.5pt}
\begin{tabular}{r|l|l|}
\cline{2-3}
& $c_0$ & $c_1$ \\
\cline{2-3}
Unaugmented & $1 - \nicefrac{1}{4} q^2 + \nicefrac{1}{2} q^3  +\ldots$ & $ -4 q^2 + 3 q^3 + \ldots$\\
Top-Level Augmented & $1 - \nicefrac{1}{2} q^3 +\ldots$ & $-4 q^2 + q^3 + \ldots$ \\
Fully Augmented & $1 $ & $ -4 q^2 + 2 q^3 +\ldots$ \\
\cline{2-3}
\end{tabular}
\caption{Fidelity coefficients for the two-level concatenated repetition code, correcting bit flip. Each fidelity is expressed as a polynomial in $p$, $F_C=\sum_k c_k p^k$, the $c_0,\,c_1$ are shown. The unaugmented code is presented, first, followed by the top-level augmented code and the fully-augmented code. }
\label{table:concatfids}
\end{table}
The tolerable initialization noise is shown in Figure \ref{fig:concat_tol_q}.
\section{Discussion \& Summary}
The large initialization errors discussed in this paper render fault-tolerant computation impossible with current methods. The purpose of the augmented error correction described above is to partially compensate for these errors, and to increase the utility of highly mixed states. This technique is intended for experimental use in the near term, in venues such as solid-state nuclear magnetic resonance (SSNMR), which does not possess an easy means of refreshing ancilla qubits, and where the error introduced by implementing the additional recovery operator is likely to be much smaller than the error in ancilla state preparation. An emphasis has been placed on avoiding the incorporation of additional ancilla qubits, since experimental implementation is currently restricted to small registers. This is true not only for SSNMR, but in other venues as well. \newline
The recovery operator in a stabilizer error correcting code, such as those shown above, is costly to implement fault-tolerantly. This is due to the fact that Pauli gates which are controlled on $n-1$ qubits are not in the Clifford group \cite{GottesmanClifford}, a set of gates that can be implemented without causing adverse error propagation in deeply concatenated error correcting codes. This has motivated the development of alternate recovery procedures, such as those used in Knill error correction~\cite{gottesman-2006-4}. It is possible that a restricted set of state preparation errors can be compensated for using Pauli gates that are controlled on a single ancilla qubit, which are in the Clifford group. It remains to be seen, however, whether this will increase the error threshold for fault-tolerant protocols.

In summary, the assumption that there exists a pure ancilla, initialized in the state $\ket{0}^{\otimes n}$ is often violated, since the initialization process is imperfect in practice. This motivates the study of error correcting codes whose encoding operators are augmented to produce higher fidelities in the presence of initialization errors. The augmentation consists of inverting the recovery operator (which performs a single-qubit unitary on the message qubit, controlled on the end state of the ancilla qubits) and inserting it before encoding. The action of this augmented encoding can be easily understood from Figure \ref{fig:genericdirtycode}; it ensures that, if the main error channel acts trivially, the output state is equal to the input state, as opposed to having been altered by the false syndrome generated by the initialization noise. This augmentation produces fidelities strictly greater than those from unaugmented codes, and constrains all error terms to be proportional to the main error channel parameter, useful when that parameter can be controlled experimentally. 
\section{Acknowledgements}
Ben Criger thanks Dr. Yaakov Weinstein for providing useful advice, and Chris Granade, for lending his expertise in parallel computing methods. The authors also thank NSERC, CIFAR, and QuantumWorks for their continued support. IQC thanks Industry Canada for their continued support.
\begin{figure}[h!]
\centering
\includegraphics[scale=0.3]{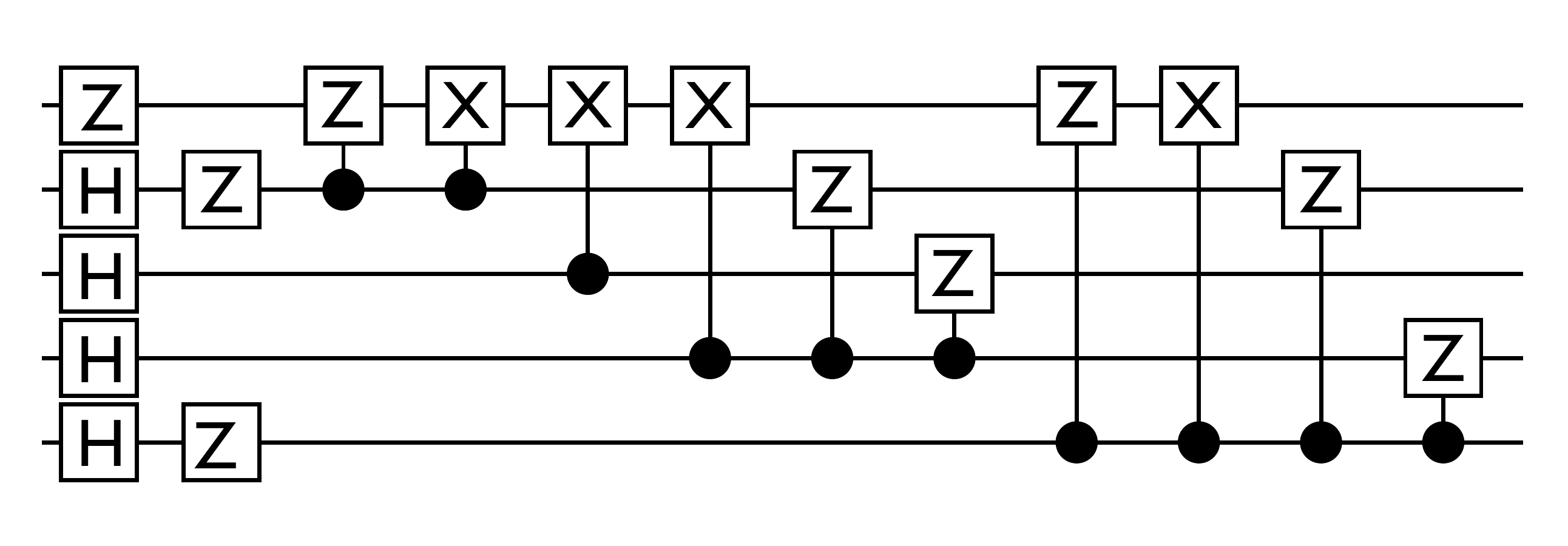}
\includegraphics[scale=0.3]{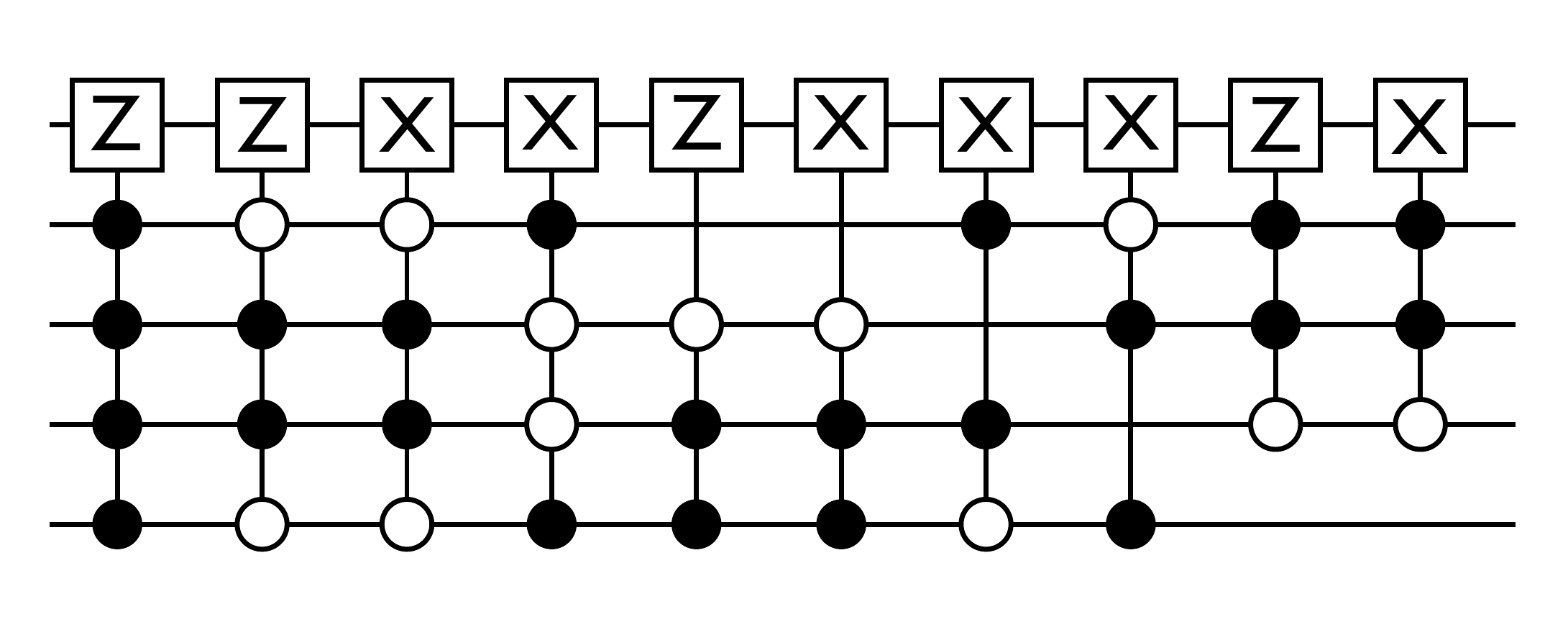}
\caption{The augmented version of the `perfect' 5-qubit code given in~\cite{gottesmanthesis} and errata. The circuit above is the unaugmented encoder, the circuit below is the correction operator. These are combined according to the prescription in Figure \ref{fig:genericdirtycode}. Here, the error channel is the depolarizing channel $\lbrace \sqrt{1-3p/4}\hat I,\,\sqrt{p/4}\hat X,\,\sqrt{p/4}\hat Y,\,\sqrt{p/4}\hat Z \rbrace$. The augmentation has a similar effect to that used on the $(2t+1)$-qubit codes countering $t^{\textrm{th}}$-order bit flip errors. We can deduce from this that the benefits of augmentation as described above are not limited to codes which counter classical errors.}
\end{figure}
\end{document}